\documentstyle[12pt,    epsfig]{article}

\def\dl{D_L}

\def\ov{\Omega_v}
\def\om{\Omega_m}
\def\b{\bigskip}

\begin{document}

\begin{center}
AN APPROXIMATE ANALYTICAL ALGORITHM FOR EVALUATING THE 
DISTANCES IN A DARK ENERGY DOMINATED 
UNIVERSE
\end{center}
\begin{center}
 T. Wickramasinghe
\\

\b

Department of Physics, The College of New Jersey, Ewing, NJ 08628
\\
Email: wick@beast.tcnj.edu

\end{center}

\b

\begin{center}
ABSTRACT
\end{center}

\b

The most recent cosmological observations indicate that the present
universe is flat and vacuum dominated. In  such a universe, the distance
measurements are always  difficult and involve numerical 
computations. 
In this paper, it is shown that the most fundamental 
distance measurement  of cosmology, the luminosity distance,   for such a universe
can be obtained in an approximate analytical way with very small
errors of less than $0.02\%$ up to %z = 5$ for any value of
vacuum energy.
The analytical calculation is shown to be
exceedingly efficient, as compared to the
traditional numerical methods.

\b

\begin{center}
{1. INTRODUCTION}
\end{center}

\b

The most recent cosmological observations indicate that the present
universe is flat and vacuum dominated such that
$\ov \approx 0.7$.
 In a such a vacuum dominated spacetime, the distance
analysis is  difficult and time consuming.
In this paper it is shown that the luminosity distance $D_L$
can be obtained to a high degree of accuracy in a purely
simple, analytical way. The analytical calculation is shown to be
exceedingly efficient and has a relative error of less than $0.02\%$ at
a redshift of $5$, as compared to numerical methods.
The analytical calculation can be performed in any
reasonable computer system and runs very fast.

\b

Current astronomical observations indicate that
the present density parameter of the universe is
$\ov + \om = 1$ and  that $\Omega_v \approx 0.7$,
where $\Omega_m$ is the contribution from all the fields
other than the  vacuum.
Thus, the calculations of distances in such a vacuum dominated
universe becomes very  important.
However, the distance calculations in such a vacuum dominated
universe involve repeated numerical
calculations and elliptic functions (Einstein 1997).
In order to simplify the numerical calculations, Pen (Pen 1999)
has developed quite an efficient
analytical recipe.

\b

In this paper, I show that a quite an elegant
analytical method, similar in many respect
to that of Pen, can be developed to
calculate the distances in a vacuum dominated
flat universe.
The analytical calculation is shown to run faster
than that of Pen and has smaller errors that
become insignificant as $z$ increases.
The paper is organized as follows.
 In \S 2,
I will develop a theory  to deduce
the luminosity distance  from
first principles.
In \S 3, I will derive necessary tools for
the analysis of  errors in the
analytical approach and will show that
the errors at $z = 1$  are of the order of  $ 0.05  \%$,
which is decreasing even further as
the redshift $z$ increases beyond unity,
making the analytical algorithm 
exceedingly efficient.

\b

\begin{center}
2. THEORY
\end{center}

\b

The most fundamental distance scale in the universe
is the luminosity distance defined by
$D_L =  \sqrt{  L / 4 \pi  F_0  }$, where $F_0$  is
the observed flux  of an astronomical
object having a luminosity
$L$.

\b

We first begin by analyzing how the scale factor
$R(t)$ varies as a function of time $t$ in a flat
 universe in which
$\ov \neq 0$.
In this case, $\dot R(t)$ is given by
(Narlikar 1983).

$$
{\dot R}^2
= H_0^2 \Omega_v R^2 + H_0^2 \Omega_m {R_0^3 \over{R}}
\eqno(1)
$$
The foregoing is immediately integrated into

$$
\left(
{R \over { R_0} }
\right)^3 =
{1 \over {2}}
{\Omega_m \over {\Omega_v}}
\left[
\cosh\left( 3 H_0 t  \sqrt{ \Omega_v}  \right) - 1 \right]
\eqno(2)
$$
Let us define
$ x  = 3 \sqrt{ \Omega_v} H_0 t $
and indicate its present value by $x_0$.
Then, Eq (2) gives

$$
x = x(z, \Omega_v) = \cosh^{-1} \left[
1 +
2
{ \Omega_v \over { 1 - \Omega_v  } }
{1 \over { (1 + z)^3 } }
\right]
\eqno(3)
$$
We note that $x$ is a monotonically decreasing
function
beyond $x_0 = x(0, 0.7) = 2.5$.
We choose the standard
Robertson-Walker metric (Peacock 1999)
as the metric of the background
spacetime.
With usual notations, this is

$$
ds^2 = c^2 dt^2 - R^2
\left[
{dr^2 \over{ 1 - k r^2 }}
+ r^2 ( d\theta^2 +
\sin^2\theta d\phi^2 )
\right]
\eqno(4)
$$
In the above spacetime, we can use Eq (2) to
obtain $r$. A simple
integration for a flat universe $(k = 0)$ yields,

$$
r  =
{c \over {H_0} R_0}
{1 \over {
3\,  \Omega_v^{1 \over 6 }  
\Omega_m^{1 \over 3 } }}
\int\limits_x^{x_0}
{ dx^{\prime}  \over {
\left[ \sinh {x^{\prime}  \over 2} \right]^{ 2 \over 3}
} }
\eqno(5)
$$
We now define a    new function
$
\Psi (x) = \lim_{\delta \rightarrow 0}
\int\limits_{\delta}^x
{ dx^{\prime}  /  {
\left[ \sinh {x^{\prime}  / 2} \right]^{ 2 \over 3}
} }
$.

\b

In the standard model  the luminosity distance
 is defined
 as
$\dl = r R_0 (1 + z)$.
Now we can  use Eq (5) to
 write the luminosity distance as

$$
\dl =  {c \over {3 H_0}}
{1 + z  \over {
  \Omega_v^{1 \over 6 }
\Omega_m^{1 \over 3 } }}
\left[
\Psi(x_0) - \Psi(x)
\right]
\eqno(6)
$$
Expanding $\Psi$ in a series expansion to the 4th order, we find that

$$
\Psi(x) =    3\;  2^{ {2 \over 3} } x^{ {1 \over 3} }
\left[ 1 - {x^2 \over 252} + {x^4 \over 21060}
\right]
+
\Psi(0)
\eqno(7)
$$
where $\Psi(0) = - 2.210$.
Now, Eq (6) reduces
to the
required expression for the luminosity distance as

$$
\dl =  {c \over {3 H_0}}
{1 + z  \over {
  \Omega_v^{1 \over 6 }
{ ( 1  - \Omega_v )  }^{1 \over 3 } }}
\left[
\Psi(x_0) - \Psi(x)
\right]
\eqno(8)
$$

\b

\vfill\eject

\begin{center}
3. CONCLUSIONS
\end{center}

\b

Eq (8) is the general form for $\dl$ in a vacuum
dominated universe.
The most popular flat model is the
the one in which $\om = \Omega_0 = 1$.
We find that our model goes very smoothly 
over to the luminosity distance in  a flat model in which
$\ov = 0$.
 
\begin{figure}[h]
\centerline{\epsfig{file=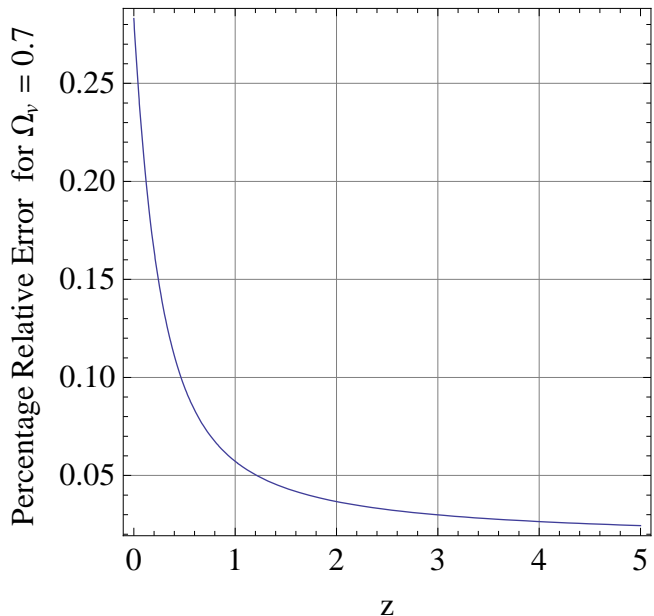}}
\caption{
Relative percentage error is plotted as a function of
$z$ for $\Omega_v = 0.7$. Notice that the error decreases sharply
up to $z = 1$ and more slowly afterwards.
 For $z=1$, the relative
error is about $0.055\%$. For $z = 5$, the error is about
$0.023\%$.
}
% \label{geometry}
\end{figure}

\b

The luminosity distance  steadily increases as the
redshift $z$.
Thus in order to get a better distance
estimate, it is important to check the
possible errors that might have crept into
the analytical approximation  in   Eq (8).

\b

The analysis was done using Eq (7).
The numerical simulations infer that there exists
a relatively small error in our method.
Our calculation shows that
the error  rapidly drops as $z$ increases.
The error at $z = 1$ is
is  estimated to be
$< 0.05 \%$.
At $z = 5$, $x = 0.207$ and  the error is about $0.023\%$.
For large $z$, the error becomes exceedingly small. 

\b

Therefore, our analytical method becomes quite   desirable as
the most interesting astronomical phenomena happen at
$z > 1$, for which the errors
in our method can safely be neglected.
Furthermore, the analytical computation is
more elegant and faster compared to
traditional numerical computations
invoked in connection
with calculations of
distances  in a vacuum dominated universe.

\b

Once we know the luminosity distance, it becomes a simple
matter to evaluate the other distances such
as the angular diameter distance and proper distance.

\b

I thank the anonymous referee for constructive suggestions
and improvements.

\b

\begin{center}
4. REFERENCES
\end{center}

\b

Einstein, D. 1997, astro-ph/9709054

Narlikar, J. V. 1983, Introduction to Cosmology,
Cambridge

Peacock, J. A. 1999, Cosmological Physics, Cambridge, 69

Pen, U. 1999, ApJ Supplement Series, 120, 49

Weinberg, S. 1972, Gravitation \&\ Cosmology, Weily

\end{document}